\documentclass[12pt]{article}
\usepackage{latexsym,epsfig}

\renewcommand{\thefootnote}{\fnsymbol{footnote}}

\begin{document}
\begin{titlepage}
\begin{flushright}
  KUNS-1790\\
  HUPD-0202\\
  UT-02-35\\
  TU-658
\end{flushright}

\begin{center}
\vspace*{10mm}
  
{\LARGE\bf Field localization in warped gauge theories}
\vspace{12mm}

{\large
Hiroyuki~Abe\footnote{E-mail address:
  abe@gauge.scphys.kyoto-u.ac.jp},
Tatsuo~Kobayashi\footnote{E-mail address:
  kobayash@gauge.scphys.kyoto-u.ac.jp},
Nobuhito~Maru\footnote{E-mail address:
  maru@hep-th.phys.s.u-tokyo.ac.jp} and
Koichi~Yoshioka\footnote{E-mail address:
  yoshioka@tuhep.phys.tohoku.ac.jp}
}
\vspace{6mm}

$^{*,\dagger}${\it Department of Physics, Kyoto University,
Kyoto 606-8502, Japan}\\[1mm]
$^*${\it Department of Physics, Hiroshima University,
Hiroshima 739-8526, Japan}\\[1mm]
$^\ddagger${\it Department of Physics, University of Tokyo,
Tokyo 113-0033, Japan}\\[1mm]
$^\S${\it Department of Physics, Tohoku University,
Sendai 980-8578, Japan}
\vspace*{15mm}

\begin{abstract}
We present four-dimensional gauge theories that describe physics on
five-dimensional curved (warped) backgrounds, which includes bulk fields 
with various spins (vectors, spinors, and scalars). Field theory on the
AdS$_5$ geometry is examined as a simple example of our
formulation. Various properties of bulk fields on this background,
e.g.,\ the mass spectrum and field localization behavior, can be
achieved within a fully four-dimensional framework. Moreover, that
gives a localization mechanism for massless vector fields. We also
consider supersymmetric cases, and show in particular that the
conditions on bulk masses imposed by supersymmetry on warped
backgrounds are derived from a four-dimensional supersymmetric theory
on the flat background. As a phenomenological application, models are
shown to generate hierarchical Yukawa couplings. Finally, we discuss
possible underlying mechanisms which dynamically realize the required
couplings to generate curved geometries.
\end{abstract}

\end{center}
\end{titlepage}

\section{Introduction}
\renewcommand{\thefootnote}{\arabic{footnote}}
\setcounter{footnote}{0}

The standard model is greatly successful but it still has many free
parameters which must be small to describe nature. While its
supersymmetric extensions, e.g.,\ the minimal supersymmetric standard
model, are attractive scenarios, small couplings are also required to
explain observed facts such as the fermion mass hierarchy and
mixing angles.

In recent years, extra dimensions have cast a new perspective on
physics beyond the standard model. One of the important aspects of extra
dimensional models is that bulk fields can be localized with
finite-width wave-function profiles. This fact provides us with a
geometrical explanation for small numbers. That is, with a
configuration where some fields are separated from each other in the
extra dimensional space, the couplings among them are generally
suppressed. Then how and where fields are localized is an issue to be
considered. From this viewpoint, extra dimensional models with a
curved background are interesting because fields could be localized
depending on the shape of the background geometry. One of the most
famous examples of curved geometries is the Randall-Sundrum (RS) model
with the AdS$_5$ warped metric~\cite{Randall:1999ee}. Field theories
of vectors, spinors, and scalars have been studied on this 
background~\cite{Goldberger:1999wh}-\cite{Gherghetta:2001kr}.
The localization behavior of zero-mode wave functions has interesting
applications to phenomenology such as the suppression of unwanted
operators. For example, hierarchical forms of Yukawa couplings and
proton decay were studied
in~\cite{Gherghetta:2000qt,Huber:2001ie}.

The localization of Kaluza-Klein (KK) excited modes also leads to
interesting phenomena. For instance, the localization of higher KK
gauge bosons could realize a composite scalar (Higgs) condensation,
which induces dynamical (electroweak) symmetry breaking on the brane
where the KK gauge bosons localize~\cite{Abe:2002yb}. In addition,
models on more complicated backgrounds where a warp factor oscillates
generate bulk fields which localize at some points in extra
dimensions~\cite{Kogan:2001wp,Dimopoulos:2001ui}. This type of
localization might be useful in explaining the observed phenomena.

However extra dimensional theories are generally nonrenormalizable
and the calculations depend on the regularization scheme that one
adopts. Furthermore, extra dimensional theories are constrained by
symmetries of higher dimensions. For example, in the supersymmetric
case, bulk theories are constrained by ${\cal N}=1$ supersymmetry in
five dimensions. Motivated by these facts, recently a 
four-dimensional (4D) description of extra dimensional models was
proposed~\cite{Arkani-Hamed:2001ca,Hill:2001mu}. With this method,
the phenomena of higher dimensional models are reproduced in terms of 4D
theories, and several interesting models have been proposed along this
line~\cite{Cheng:2001vd,Csaki:2002em}.

In this paper, we present 4D gauge theories that describe physics on
5D curved geometries. As will be discussed below, taking generic values
of gauge couplings and gauge-symmetry-breaking vacuum expectation
values (VEVs), the models provide vector, spinor, and scalar fields on
curved extra dimensions.\footnote{In the same spirit, curved backgrounds 
were studied in \cite{Sfetsos:2001qb}.} 
As a good and simple illustration, we compare our 4D model with the RS
one. We particularly focus on the ``localization'' behaviors of mass
eigenstates in ``index spaces'' of gauge groups. It will be shown that
the localization profiles and the exponentially suppressed massive spectrum
are certainly reproduced. In addition, our formulation gives a
localization mechanism even for massless vector fields. As a
phenomenological application, hierarchical Yukawa matrices are derived
in our approach; that is a hierarchy without symmetries in four dimensions.

The localization behavior depends on the required conditions for 
gauge-symmetry-breaking VEVs and gauge and other couplings. If these values 
are determined in the underlying theories, it may be said that the physics on
warped backgrounds is dynamically generated within a four-dimensional
framework. We consider several possibilities to realize the conditions
by utilizing, for example, strongly coupled gauge theories. Thus this could
provide a purely 4D dynamical approach for small numbers.

We will proceed with the argument as follows. In Sec.~\ref{sec:DWD},
we describe our 4D gauge theories, which have generic (nonuniversal)
values of gauge-symmetry-breaking VEVs and couplings. The models
provide vector, spinor, and scalar fields in warped extra
dimensions. It is also shown that supersymmetry multiplets in flat 4D
models generate supersymmetry multiplets on warped backgrounds. In
Sec.~\ref{sec:NE}, we then numerically determine with a finite number of
gauge groups that the formulation given in Sec.~2 certainly reproduces
various properties of bulk fields on the RS background. In addition, a
phenomenological application to quark mass matrices is also
given. Finally, we discuss possibilities of dynamically realizing the
conditions required for curved geometries in Sec.~\ref{sec:DDWD}. We
conclude the discussion in Sec.~\ref{sec:conclusion}. The Appendix is
devoted to a brief review of 5D bulk fields on a RS background.

\section{4D formulation for curved geometries}
\label{sec:DWD}

\subsection{Vectors}

\begin{figure}[t]
\centerline{\epsfig{figure=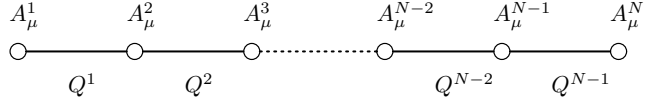,width=0.5\linewidth}}
\caption{Moose diagram for bulk vector fields
in the orbifold extra dimension.}
\label{fig:links}
\end{figure}

Following Refs.~\cite{Arkani-Hamed:2001ca,Hill:2001mu}, we introduce
${\rm SU}(n)_i$ gauge theories with gauge couplings $g_i$ ($i=1,\ldots,N$),
and scalar fields $Q_i$ [$i=1,\ldots,(N-1)$] which are in
bifundamental representations of ${\rm SU}(n)_i\times {\rm SU}(n)_{i+1}$.
The system is schematized by the segment diagram in Fig.~\ref{fig:links}.

The gauge invariant kinetic term of the scalars $Q_i$ is written by
\begin{eqnarray}
  {\cal L} &=& \sum_{i=1}^{N-1} (D_\mu Q_i)^\dagger (D^\mu Q_i),
\label{eq:Qkin}
\end{eqnarray}
where the covariant derivative is given by
$D_\mu Q_i=\partial_\mu Q_i-ig_iA^i_\mu Q_i+ig_{i+1}Q_i A^{i+1}_\mu$.
We assume that the scalar fields $Q_i$ develop VEVs proportional
to the unit matrix, $\langle Q_i \rangle = v_i$, which break the gauge
symmetry to a diagonal ${\rm SU}(n)_{\rm diag}$. From the kinetic term
(\ref{eq:Qkin}), the mass terms for the vector fields $A_\mu^i$ are 
obtained: 
\begin{eqnarray}
  {\cal L}_{gm} &=& \frac{1}{2}\sum_{i=1}^{N-1}
  \Big|v_i(g_{i+1}A_\mu^{i+1}-g_iA_\mu^i)\Big|^2 \label{gKT}
  \ = \ \frac{1}{2} \sum_{i=1}^{N-1}\sum_{j=1}^N
  \left|D_{ij}^{\,\rm vec} A_\mu^j\right|^2,
\label{eq:gau-mass}
\end{eqnarray}
where the $(N-1)\times N$ matrix $D^{\,\rm vec}$ is defined as
\begin{eqnarray}
  D^{\,\rm vec} &=& \pmatrix{ v_1 & & \cr & \ddots & \cr & & v_{N-1} }
  \pmatrix{ -1 & 1 & & \cr
    & \ddots & \ddots & \cr
    & & -1 & 1 }
  \pmatrix{ g_1 & & \cr & \ddots & \cr & & g_N }.
\label{eq:diffop}
\end{eqnarray}

The consequence of these mass terms is that we have a massless gauge boson
corresponding to the unbroken gauge symmetry, which is given by the
following linear combination:
\begin{equation}
  \tilde A^{(0)}_\mu \;=\; \sum^N_{i=1}
  \biggl(\frac{g_{\rm diag}}{g_i}\biggr) A^i_\mu,
\label{zero-A}
\end{equation}
where $g_{\rm diag}^{-2}\equiv\sum_{i=1}^N g_i^{-2}$ and $g_{\rm diag}$ 
is the gauge coupling of the low-energy gauge theory ${\rm SU}(n)_{\rm diag}$. 
The profile of $\tilde A_\mu^{(0)}$ is independent of the values 
of $v_i$. It is found from this that the massless vector field is 
``localized'' at the points with smaller gauge couplings. If the gauge
couplings take a universal value $^\forall g_i=g$, the massless 
mode $\tilde A^{(0)}_\mu$ has a constant ``wave function'' along 
the ``index space'' of gauge groups. As seen below, this direction
labeled by $i$ becomes the fifth spatial dimension in the continuum 
limit ($N\to\infty$). The localization behavior can easily be
understood from the fact that, for smaller gauge coupling $g_i$,
the symmetry-breaking scale $g_iv$ of ${\rm SU}(n)_i$ becomes lower,
and hence the corresponding vector field $A_\mu^i$ becomes the more 
dominant component in the low-energy degree of
freedom $\tilde A^{(0)}_\mu$.

It is interesting to note that this vector localization mechanism 
ensures charge universality. Suppose that there is a field in a 
nontrivial representation of ${\rm SU}(n)_i$ only. That is, it couples 
only to $A^i_\mu$ with strength $g_i$. This corresponds to a 
four-dimensional field confined on a brane. If there are several 
such fields, they generally have different values of gauge 
couplings. However, note that these fields couple to the massless 
modes $\tilde A_\mu^{(0)}$ with a {\it universal} gauge 
coupling $g_{\rm diag}$ defined above. This is because, in the presented 
mechanism, the vector fields are localized depending on the values of the 
gauge couplings. 

\medskip

As for massless eigenstates, the mass eigenvalues and wave functions 
are obtained by diagonalizing the mass matrix (\ref{eq:diffop}). 
The simplest case is the universal couplings 
\begin{equation} 
^\forall v_i \,=\, v, \qquad ^\forall g_i \,=\, g. 
\end{equation}
In this case, one obtains the mass eigenvalues of $D^{\,\rm vec}$ as 
\begin{eqnarray}
m_n &=& 2gv \sin\biggl(\frac{n\pi}{2N}\biggr) \qquad
(n=0,\ldots,N-1). 
\end{eqnarray}
In the limit $N\to\infty$ with $L\equiv 2N/gv\,$ fixed (the limit to 
continuum 5D theory), the eigenvalues become 
\begin{eqnarray}
m_n \,\simeq\, \frac{2n\pi}{L}. 
\end{eqnarray}
These are the same spectrum as that of the bulk gauge boson 
in the $S^1/Z_2$ extra dimension with radius $L/2\pi$. 

With generic values of VEVs $v_i$ and gauge couplings $g_i$, the 
situation is rather complicated. In this case, the mass term 
(\ref{eq:gau-mass}) of the vector fields becomes 
\begin{eqnarray}
  {\cal L}_{gm} &=& \frac{1}{2}\sum_i v_i^2g_ig_{i+1}
  (A_\mu^{i+1}-A_\mu^i)^2 \nonumber\\
  && -\frac{1}{2}\sum_i \Bigl[v_i^2g_{i+1}(g_{i+2}-2g_{i+1}+g_i)
  +(v_{i+1}^2-v_i^2) g_{i+1}(g_{i+2}-g_{i+1})\Bigr](A_\mu^{i+1})^2
  \nonumber \\[1mm]
  && +\frac{1}{2}v_N^2g_N(g_{N+1}-g_N)(A_\mu^N)^2
  -\frac{1}{2} v_1^2g_1(g_2-g_1)(A_\mu^1)^2. 
  \label{eq:gbbm}
\end{eqnarray}
The first term becomes the kinetic energy transverse to the 
four dimensions in the continuum limit. On the other hand, the second 
and third terms are bulk and brane mass terms, respectively. It should 
be noted that these mass terms vanish in the case of universal gauge 
coupling, which corresponds to a flat massless vector field in 5D 
theory as discussed above. In other words, nonuniversal gauge 
couplings generate bulk/brane mass terms and cause a localization of 
the wave function.

\subsubsection{VEVs and couplings generating AdS$_5$ background}
First we consider the series of VEVs $v_i$ and couplings $g_i$ 
that generates a vector field on the RS warped background, 
namely, the AdS$_5$ background. 
This model can be obtained by choosing a universal $g_i$ 
and by varying $v_i$ as 
\begin{eqnarray}
\mbox{RS} : \qquad  g_i \,=\, g, \qquad
v_i \,=\, ve^{-ki/(gv)}.
\label{eq:RSparaf}
\end{eqnarray}
Substituting this and taking the continuum limit, Eq.~(\ref{gKT})
becomes
\begin{eqnarray}
  {\cal L}_{gm} &=& \frac{1}{2} \int_0^{L/2}\! dy\,
  \Bigl[e^{-ky}\partial_y A_\mu(x,y)]^2,
\label{eq:contlimrs}
\end{eqnarray}
where $y$ represents the coordinate of the extra dimension:
$y \leftrightarrow i/gv$ ($i=1,\ldots,N$) and
$A_\mu^i(x) \leftrightarrow A_\mu(x,y)$, etc. It is found that
Eq.~(\ref{eq:contlimrs}) successfully induces the kinetic energy term 
along the extra dimension and mass terms for the vector field on the 
warped background metric
\begin{equation}
  ds^2 = G_{MN}dx^M dx^N = e^{-2ky} \eta_{\mu\nu}dx^\mu dx^\nu-dy^2, 
\label{eq:rsbkg}
\end{equation} 
where $\eta_{\mu\nu}={\rm diag}(1,-1,-1,-1)$ with $\mu=0,1,2,3$.
We here conclude that we can obtain the vector field on a RS warped 
background by varying only the VEVs $v_i$. In the following we will 
see that nonuniversal gauge couplings $g_i$ induce other interesting 
results beyond the effects from the background metric. 

\subsubsection{Abelian case with nonuniversal gauge couplings}
Now let us compare the 4D model with generic couplings (\ref{eq:gbbm}) 
to extra dimensional ones. We define the dimensionless parameters $f_i$
and $h_i$ as
\begin{eqnarray}
  g_i \,=\, gf_i, \qquad v_i \,=\, vh_i.
\label{eq:deffh}
\end{eqnarray}
First we restrict ourselves to the case that the gauge group 
is ${\rm U}(1)$, namely, Abelian theory with {\it no vector self-couplings}. 
Similarly substituting Eq.~(\ref{eq:deffh}) and taking the continuum limit, 
Eq.~(\ref{gKT}) becomes 
\begin{eqnarray}
  {\cal L}_{gm} &=& \frac{1}{2} \int_0^{L/2}\! dy\,
  \Bigl[h(y)\partial_y\bigl[f(y)A_\mu(x,y)\bigr]\Bigr]^2. 
\label{eq:contlim}
\end{eqnarray}
Equation (\ref{eq:contlim}) induces the kinetic energy term along the extra 
dimension and mass terms for the vector field on the warped background 
metric: 
\begin{equation}
  ds^2 = G_{MN}dx^M dx^N = [f(y)h(y)]^2 \eta_{\mu\nu}dx^\mu dx^\nu-dy^2. 
\label{eq:generalbkg}
\end{equation} 
The bulk and boundary mass terms are $y$ dependent and proportional to 
the derivatives of $f(y)$. This is also seen from the 4D model 
[the second and third terms in Eq.~(\ref{eq:gbbm})]. 

The above is a generic correspondence between our 4D case and 
continuum 5D theory. As an example, consider the following forms of 
the parameters: 
\begin{eqnarray}
  f(y) \,=\, e^{-\zeta ky}, \qquad h(y) \,=\, e^{-\eta ky},
\end{eqnarray}
where $k$ is a positive constant with mass dimension $1$. 
Equation (\ref{eq:contlim}) leads to 
\begin{eqnarray}
  {\cal L}_{gm} &=& \frac{1}{2} \int_0^{L/2}\! dy\,
  e^{-2(\zeta+\eta)ky} \biggl[(\partial_y A_\mu)^2
  -\zeta(2\eta+\zeta)k^2 (A_\mu)^2 \biggr]
  -\frac{1}{2}\biggl[\zeta ke^{-2(\zeta+\eta)ky}
  (A_\mu)^2\biggr]_0^{L/2}. 
\label{eq:wgmbbm}
\end{eqnarray}
The first term on the right hand side is the kinetic term of 
the gauge boson along the extra dimension with the warped background 
\begin{equation}
ds^2 \,=\, e^{-2(\zeta +\eta)ky}\eta_{\mu\nu}dx^\mu dx^\nu-dy^2. 
\end{equation}
The second and third terms correspond to the bulk and boundary 
masses announced before. As easily seen, the above equation includes 
the expression for vector fields on the RS background. In the 5D 
RS model, the Lagrangian for vector fields is written as (see the Appendix) 
\begin{eqnarray}
  {\cal L}_{\rm RS}^{\,\rm vec} &=& -\frac{1}{4}F_{\mu\nu}F^{\mu\nu}
  +\frac{1}{2} e^{-2ky}(\partial_y A_\mu)^2,
\label{RSgauge}
\end{eqnarray}
where the $A_5=0$ gauge fixing condition is chosen. By comparing
Eq.~(\ref{eq:wgmbbm}) with Eq.~(\ref{RSgauge}), we find that the case
with
\begin{equation}
\zeta+\eta \,=\, 1
\end{equation}
realizes vector fields on the RS background. Also, a special 
limit, $(\zeta,\eta)=(0,1)$, produces the flat zero-mode solution. 
That corresponds to the form of the parameters (\ref{eq:RSparaf}) 
in the previous special argument. 
The other solutions which satisfy $\zeta+\eta=1$ correspond to 
nonflat wave functions of the zero-mode vector field on the RS 
background. It is clearly understood in our formulation that such 
nonflat wave functions are caused by introducing bulk and/or 
boundary mass terms in the RS model. For example, in the case 
of $(\zeta,\eta)=(1,0)$, the vector field has bulk and boundary mass 
terms, and is localized with a peak at the $y=L/2$ point. It should be 
noticed that with these bare mass terms the zero mode is still 
massless. This is understood from our formulation where the gauge 
symmetry $SU(n)_{diag}$ is left unbroken in the low-energy effective 
theory. 

\subsubsection{Non-Abelian case with nonuniversal gauge couplings}
In the above Abelian case we discussed interpretation of the nonuniversal 
$f_i$ as $y$-dependent bulk or boundary masses in the 
warped extra dimension. Next we treat the non-Abelian theory with 
vector self-couplings. Since in this case we also have $y$-dependent
vector self-couplings in addition to the $y$-dependent bulk or boundary 
masses, it may be convenient and instructive to see $f_i$ 
as a $y$-dependent coefficient of the vector kinetic term.
To this end, we define the four-dimensional field $A_5^i$,
\begin{eqnarray}
Q_i \equiv v_i e^{-ia (g_i A_5^i + g_{i+1} A_5^{i+1})/2} 
    = v_i \left(1-ia (g_i A_5^i + g_{i+1} A_5^{i+1})/2
                 +{\cal O}(a^2)\right),
\label{eq:defa5}
\end{eqnarray}
where $a\equiv L/(2N)$ is the lattice spacing, which goes to zero in
the continuum limit. 
Rescaling the gauge fields 
$\sqrt{N} f_i (A^i_\mu,\,A^i_5) \to (A^i_\mu,\,A^i_5)$,
the kinetic term 
$-\frac{1}{4} \sum_{i=1}^N F_{\mu \nu}^i F^{i \mu \nu}$
and Eq.~(\ref{eq:Qkin}) become 
\begin{eqnarray}
  {\cal L}^{gQ}_{\rm kin} 
&=& -\frac{1}{4} \sum_{i=1}^N a \frac{f^{-2}_i}{L/2} 
  F^i_{\mu \nu} F^{i\mu \nu} \nonumber \\ &&
    +\sum_{i=1}^{N-1} a \frac{h^2_i}{L/2}
    \Bigg| \partial_\mu A_5^{i+1/2} - \frac{A^{i+1}_\mu-A^i_\mu}{a} 
           -i \hat{g} [A^i_\mu, A_5^{i+1/2}] 
\nonumber \\ && \qquad \qquad \qquad \qquad  \qquad \qquad  
           +i \hat{g} A_5^{i+1/2} (A^{i+1}_\mu - A^i_\mu) 
           + {\cal O}(a^{1/2}) \Bigg|^2, 
\label{eq:gQkin2}
\end{eqnarray}
where $\hat{g}=g/\sqrt{N}$
and $A_5^{i+1/2}\equiv (A_5^i+A_5^{i+1})/2$. 
In the continuum limit  
$N \to \infty$ with $L$ and $\hat{g}$ fixed,
Eq.~(\ref{eq:gQkin2}) results in 
\begin{eqnarray}
  {\cal L}^{gQ}_{\rm kin} 
&=& -\frac{1}{4} \int_{0}^{L/2}dy \frac{f^{-2}(y)}{L/2}
      \Big\{  \eta^{\mu \nu} \eta^{\rho \sigma} 
             F_{\mu \rho} F_{\nu \sigma}
            -2[h(y)f(y)]^2 \eta^{\mu \nu} F_{\mu 5} F_{\nu 5} 
      \Big\},
\end{eqnarray}
where $F_{MN}(x,y) = 
\partial_M A_N (x,y) - \partial_N A_M (x,y) 
- - i\hat{g} [A_M (x,y), A_N (x,y)]$. 
This completely reproduces the 5D Yang-Mills 
kinetic term with a $y$-dependent coefficient
\begin{eqnarray}
  {\cal L}^{\rm vec}_{\rm warp} 
&=& -\frac{1}{4} \int_0^{L/2}\! dy\, f^{-2}(y) 
  \sqrt{-G} G^{MN}G^{AB} F_{MA}F_{NB}
\label{eq:5dym}
\end{eqnarray}
on the warped-background metric (\ref{eq:generalbkg}),
provided that $g_{\rm 5D} = \sqrt{L/2}\,\hat{g}$. 
This is the generic correspondence between 
the present 4D model and continuum 5D theory. From 
Eq.~(\ref{eq:5dym}), we thus find the $y$-dependent factor 
$f^{-2}(y)$ in front of the canonical Yang-Mills term, which
corresponds to a 5D dilaton VEV\@.
The factor does carry the origin of the massless
vector localization shown in Eq.~(\ref{zero-A}).\footnote{For a
continuum 5D analysis, see \cite{Kehagias:2000au}.} 
With the constant gauge coupling $g_i \equiv g$ ($f_i$ = 1), one
obtains a bulk vector field with a constant zero mode on the 
warped metric (\ref{eq:generalbkg}).
A field redefinition $f^{-1}(y) A_M \to A_M$ in Eq.~(\ref{eq:5dym}) 
gives the previous bulk and boundary mass terms 
but one then has $y$-dependent vector self-couplings in non-Abelian cases. 

\subsection{Spinors}
\label{sec:spinor}

We next consider spinor fields by arranging fermions of fundamental or 
antifundamental representation in each gauge theory ${\rm SU}(n)_i$. We 
introduce two Weyl (one Dirac) spinors to construct a 5D bulk 
fermion. The orbifold compactification in continuum theory requires that 
one spinor obeys the Neumann boundary condition and the other the 
Dirichlet one. In the present 4D model, this can be achieved by 
having asymmetrical numbers of fundamental and antifundamental spinors, 
resulting in chiral fermions in the low-energy gauge theory. Here we 
consider the fundamental Weyl spinors $\eta_i$ ($i=1,\ldots,N$) in 
the ${\rm SU}(n)_i$ theory and the 
antifundamental $\psi_j$ ($j=1,\ldots, N-1$). As seen 
below, $\eta$ corresponds to the bulk fermion with the Neumann boundary
condition and $\psi$ to that with the Dirichlet one.

The generic gauge-invariant mass and the mixing terms of $\eta_i$
and $\psi_j$ are written as
\begin{eqnarray}
{\cal L}_{fm} &=& -\sum_{i=1}^{N-1}(\alpha_i \psi_i Q_i \eta_{i+1}
-\beta_iv_i \psi_i \eta_i) +\textrm{H.c.},
\label{eq:fermionmoose}
\end{eqnarray}
where $\alpha_i$ and $\beta_i$ are dimensionless coupling 
constants. We assume that $Q_i$ develop 
VEVs $\langle Q_i \rangle=v_i$. The mass matrix for spinors is then
given by
\begin{equation}
D^{\,\rm spi} \;=\; \pmatrix{ v_1 & & \cr & \ddots & \cr & & v_{N-1} }
  \pmatrix{ -\beta_1 & \alpha_1 & & \cr
    & \ddots & \ddots & \cr
    & & -\beta_{N-1} & \alpha_{N-1} }.
\label{eq:mass-spi}
\end{equation}
The spinor mass eigenvalues and eigenvectors (wave functions) are
read from this matrix. One easily sees that the massless mode is
contained in $\eta$ and given by the following linear combination:
\begin{equation}
\tilde \eta^{(0)} = \frac{1}{\sqrt{\sum_{i=1}^N
  \Bigl(\prod_{j=1}^{i-1}\frac{\beta_j}{\alpha_j}\Bigr)^2}}
  \sum_{i=1}^N \biggl(
  \prod_{j=1}^{i-1}\frac{\beta_j}{\alpha_j}\biggr)\,\eta_i.
\label{zero-eta}
\end{equation}
Therefore the localization profile of zero mode depends on the
ratio of dimensionless couplings $\alpha_i$ and $\beta_i$. A simple
case is $\alpha_i=\beta_i$ for all $i$. In this case,
$\tilde \eta^{(0)}$ corresponds to a chiral zero mode obtained from
a 5D bulk fermion on the flat background. If $\alpha_i\neq\beta_i$,
the system describes a fermion with a curved wave-function profile.
For example, if $\beta_i/\alpha_i=c>1$ ($<1$), $\tilde \eta^{(0)}$
has a monotonically increasing (decreasing) wave-function profile.
As another interesting example, 
taking $\beta_i/\alpha_i=c\,x^i$ ($c$, $x$ are constants and $x <1$), 
$\tilde \eta^{(0)}$ has a Gaussian profile with a peak 
at $i = 1/2-\ln_x c$. Other profiles of massless chiral fermions could
also be realized in our approach.

\medskip

Let us discuss the 5D continuum limit. The relevant choice of 
couplings $\alpha_i$ and $\beta_i$ is
\begin{eqnarray}
\alpha_i =g_{i+1}, \qquad
\beta_i =g_i(1-c_i).
\label{eq:spinorcond}
\end{eqnarray}
The parameters $c_i$ give rise to a bulk bare mass in the continuum
limit as will be seen below. The only difference between the vector and
spinor cases is the existence of possible bulk mass parameters 
[see Eqs.~(\ref{eq:diffop}) and (\ref{eq:mass-spi})]. 
The mass and mixing terms~(\ref{eq:fermionmoose}) then become
\begin{eqnarray}
{\cal L}_{fm} &=& -\int_0^{L/2}dy\,h(y)
\biggl[ \psi(x,y)\partial_y \Bigl\{f(y) \eta(x,y)\Bigr\}
+gvf(y)c(y) \psi(x,y) \eta(x,y) \biggr] + \textrm{H.c.},
\end{eqnarray}
where $f$ and $h$ are the same as defined in the case of vector
fields~(\ref{eq:deffh}). Similar to the vector case, this form is
compared with the bulk spinor Lagrangian in the RS space-time 
(see the Appendix)
\begin{eqnarray}
{\cal L}_{\rm RS}^{\,\rm spi}
&=& -\int_0^{L/2}dy\, \Bigg[
\bar\psi i \bar\sigma^\mu \partial_\mu \psi
+\bar\eta i \bar\sigma^\mu \partial_\mu \eta
+\left\{ \psi e^{-ky} \Big( \partial_y + (c-1/2) \sigma' \Big)
  \eta + \textrm{H.c.} \right\} \Bigg].
\label{eq:RSspinor}
\end{eqnarray}
Here the kinetic terms have been canonically normalized in order to
compare them to the 4D model. In Eq.~(\ref{eq:RSspinor}), $c$ is a
possible 5D Dirac mass, and the ``1/2'' contribution in the mass terms
comes from the spin connection with the RS metric. It is interesting
that the 5D spinor Lagrangian (\ref{eq:RSspinor}) is reproduced by
taking the exact same limit of parameters as that in the vector
case, defined by Eq.~(\ref{eq:RSparaf}). Furthermore, the relation
between the mass parameters $c$ should be taken as
\begin{eqnarray}
c_i \;=\; \left(c-\frac{1}{2} \right)\frac{k}{gv}.
\label{eq:ci-spi}
\end{eqnarray}
That is, the $c_i$'s take a universal value. Now the localization 
behavior of the spinors is easily understood. In the present 4D model, the 
spinor mass matrix (\ref{eq:mass-spi}) becomes with 
Eq.~(\ref{eq:spinorcond}) 
\begin{equation}
D^{\,\rm spi} \;=\; \pmatrix{ v_1 & & \cr & \ddots & \cr & & v_{N-1} }
\pmatrix{ -1+c_1 & 1 & & \cr
  & \ddots & \ddots & \cr
  & & -1+c_{N-1} & 1}
\pmatrix{ g_1 & & \cr & \ddots & \cr & & g_N }.
\label{eq:mass-RSspi}
\end{equation}
A vanishing bulk mass parameter $c=1/2$ corresponds to $^\forall c_i=0$,
that is, $\alpha_i=\beta_{i+1}$ in our model. Then the mass 
matrix $D^{\,\rm spi}$ is exactly the same as $D^{\,\rm vec}$, and 
their mass eigenvalues and eigenstates are the same. In particular, 
the massless mode $\tilde \eta^{(0)}$ has a flat wave function with 
universal gauge couplings as considered here. This is consistent with 
the expression (\ref{zero-eta}), where the ratio $\alpha_i/\beta_i$
determines the wave-function profile. On the other hand, in the case 
of $c>1/2$ ($c<1/2$), the RS zero-mode spinor is localized 
at $y=0$ ($y=L/2$)~\cite{Goldberger:1999wh}, which in turn corresponds
to $c_i>0$ ($c_i<0$) in our model. One can see from the spinor mass
matrix (\ref{eq:mass-RSspi}) that the zero-mode wave function is
monotonically increasing (decreasing) with respect to the index $i$.

In this way, we have a 4D localization mechanism for the spinor fields. 
Nonuniversal gauge couplings or nonuniversal masses give rise to 
a nonflat wave function for a chiral massless fermion. The latter 
option is not realized for vector fields. Notice that the charge
universality still holds in the low-energy effective theory. That is,
with any complicated wave-function profiles, zero modes interact with 
a universal value of the gauge coupling. This is because curved profiles
of vector fields depend only on the gauge couplings.

\subsection{Scalar}
\label{sec:scalar}

Finally we consider scalar fields. We may introduce two types of
scalar field $\phi_i$ and $\varphi_i$ in the fundamental and
antifundamental representations of ${\rm SU}(n)_i$, respectively. In
addition, for each type of scalar, there are two choices of the $Z_2$
parity assignment in the continuum limit. This orbifolding procedure
is incorporated by removing $\phi_1$ or $\varphi_N$. The gauge
invariant mass and mixing terms for $\phi$ and $\varphi$ are written as
\begin{eqnarray}
  {\cal L}_{sm}^\phi
  &=& -\sum_{i=1}^N \Bigl|\alpha'_iQ_i\phi_{i+1}
  -\beta'_iv_i\phi_i\Bigr|^2
  -\sum_{i=1}^N (g_iv_i\gamma_i)^2 |\phi_i|^2,
  \label{eq:nsp} \\
  {\cal L}_{sm}^\varphi
  &=& -\sum_{i=1}^N \Bigl|\bar\alpha'_iQ_i\varphi_i
  -\bar\beta'_iv_{i+1}\varphi_{i+1}\Bigr|^2
  -\sum_{i=1}^N (g_iv_i\bar\gamma_i)^2 |\varphi_i|^2,
  \label{eq:dsp}
\end{eqnarray}
where $\alpha$, $\beta$, and $\gamma$ are the dimensionless coupling 
constants. It is implicitly assumed that nonintroduced fields are 
appropriately removed in the sums. We have included the mixing mass 
terms up to the nearest-neighbor interactions. Other invariant terms 
such as $|\phi_i\phi_j|^2$ or terms containing $Q_i$ correspond to 
nonlocal interactions in 5D theories, and we do not consider these in 
this paper. Notice, however, that for a supersymmetric case, these 
terms may be suppressed due to renormalizability and holomorphy of 
the superpotential. 

The zero-mode eigenstates are given in the same form as that of the spinor 
shown in the previous section, replacing $\alpha$ and $\beta$ by 
$\alpha'$ and $\beta'$ ($\bar\alpha'$ and $\bar\beta'$). Therefore the
ratio $\alpha'/\beta'$ ($\bar\alpha'/\bar\beta'$) determines the
zero-mode wave function.

\medskip

Let us consider the continuum 5D limit. In what follows, we 
remove $\varphi_N$, which corresponds to the $Z_2$ assignment $\phi(+)$ 
and $\varphi(-)$. The 5D limit can be achieved by taking the following
choices of couplings:
\begin{eqnarray}
&& \alpha'_i \,=\, g_{i+1}, \qquad \beta'_i \,=\, g_i(1-c'_i), \\[1mm]
&& \bar\alpha'_i \,=\, g_{i+1}, \qquad
\bar\beta'_i \,=\, g_{i+1}(1-\bar c'_{i+1}),
\end{eqnarray}
where $c'_i$ and $\bar c'_i$ correspond to the bulk mass parameters, as in 
the spinor case. Then the mass terms (\ref{eq:nsp}) and (\ref{eq:dsp}) 
for the scalars take the following forms with the parametrization 
(\ref{eq:deffh}): 
\begin{equation}
{\cal L}_{sm}^\phi
\,=\, -\int_0^{L/2} dy\, h^2(y)\biggl[\,
\Bigl|\bigl[\partial_y +gvc'(y)\bigr]\{f(y)\phi(x,y)\}\Bigr|^2
+[gv\gamma(y)]^2|f(y)\phi(x,y)|^2 \biggr],
\end{equation}
\begin{equation}
{\cal L}_{sm}^\varphi
\,=\, -\int_0^{L/2} dy\, f^2(y)\biggl[\,
\Bigl|\bigl[\partial_y -gv\bar c'(y)\bigr]\{h(y)\varphi(x,y)\} \Bigr|^2
+[gv\bar\gamma(y)]^2|h(y)\varphi(x,y)|^2 \biggr].
\end{equation}

As a special case, we compare $\phi$ and $\varphi$ with the scalar
fields in the RS space-time. The scalar Lagrangian on the RS
background is (see also the Appendix)
\begin{eqnarray}
{\cal L}_{\rm RS}^{\,\rm sca}
&=& -\int_0^{L/2} dy\, \left[ |\partial_\mu\phi|^2 +\left|e^{-ky}
    \left\{\partial_y+(1-b)k\right\} \phi\right|^2
  +(a+4b-b^2) k^2 e^{-2ky} |\phi|^2 \right]
\end{eqnarray}
where the 4D kinetic term is canonically normalized, and $a$ and $b$ 
are the bulk and boundary mass parameters, respectively, defined in the 
Appendix [Eq.~(\ref{eq:smass})]. By substituting the RS limit in our 
model given by Eq.~(\ref{eq:RSparaf}), we find the relations between the 
mass parameters in 4D and 5D: 
\begin{eqnarray}
  c'_i \,=\, (1-b)\frac{k}{gv}, \qquad
  \gamma_i^2 \,=\, (a+4b-b^2) \biggl(\frac{k}{gv}\biggr)^2
\qquad \textrm{ (for $\phi$)},
\label{eq:massn} \\
  \bar c'_i \,=\, (b-2)\frac{k}{gv}, \qquad
  \bar\gamma_i^2 \,=\, (a+4b-b^2)\biggl(\frac{k}{gv}\biggr)^2
\qquad \textrm{ (for $\varphi$)}.
\label{eq:massd}
\end{eqnarray}

\subsection{Supersymmetry on warped background}

In this subsection, we discuss 5D supersymmetry on warped 
backgrounds. Generally a supersymmetric theory may be obtained by 
relevant choices of couplings from a nonsupersymmetric theory. We 
here examine whether it is possible to construct supersymmetric 4D 
models which describe 5D supersymmetric ones on warped 
backgrounds. This is a nontrivial check for the ability of our formalism 
to properly describe 5D nature. In Ref.~\cite{Gherghetta:2000qt}, 
the 5D theory on the AdS$_5$ RS background was studied. 
There, supersymmetry on AdS$_5$ geometry was identified and then the 
conditions on the mass parameters imposed by this type of supersymmetry 
were derived (also given in the Appendix here).
As seen below, these relations among mass parameters for AdS$_5$
supersymmetry are indeed satisfied in our {\it 4D formalism}. This 
fact seems remarkable in the sense that the present analyses do not
include gravity.

First consider vector supermultiplets in 5D. The scalar fields $Q_i$
and the gauge bosons $A^i_\mu$ are extended to chiral and vector
superfields in 4D, respectively. Notice that the VEVs that were 
discussed above,
\begin{equation}
(Q_i)^a_b \,=\, v_i\,\delta^a_b,
\end{equation}
are in the (baryonic) $D$-flat direction.

We start with the following 4D supersymmetric Lagrangian
\begin{equation}
{\cal L} \;=\; \sum^{N-1}_{i=1} \int d^2\theta d^2 \bar \theta\,
Q^\dagger_i e^{\sum V}Q_i +
\sum^{N}_{i=1}{1 \over 4g^2_i} \int d^2\theta W^i W^i
+ \textrm{H.c.}.
\end{equation}
The bilinear terms of the component fields are written in the unitary gauge
(we follow the conventions of \cite{WB}): 
\begin{eqnarray}
&& \frac{1}{2}(D_{ij} A^j_\mu)^{\rm t}(D_{ik} A^k_\mu)
+\frac{1}{2}(D^i)^2 +\frac{1}{2}(D_{ij}N^j)^\dagger (D_{ik}N^k)
\nonumber \\ && \hspace{1cm}
+(D_{ij}^{\rm t} C^j)^{\rm t}(D_{ik} D^k)
-(D_{ij} \chi^j)(D_{ik} \lambda^k)
\;\;\textrm{+h.c}. ,
\end{eqnarray}
where we have rescaled
$(A^i_\mu,\lambda^i,D^i) \rightarrow g_i(A^i_\mu,\lambda^i,D^i)$ for
canonical normalization of the kinetic terms. The mass matrix $D_{ij}$ is
defined in Eq.~(\ref{gKT}). The first term is nothing but
Eq.~(\ref{gKT}), that is, the mass terms for vector fields. By also
canonically normalizing $C^i$ and $\chi^i$ and integrating out the
auxiliary fields, we find the mass terms for the adjoint spinors and
scalar
\begin{equation}
-\chi^i D_{ij} \lambda^j +\textrm{H.c.}
-\frac{1}{2}(D^{\rm t}_{ij} C^j)^{\rm t}(D^{\rm t}_{ik} C^k).
\end{equation}
These masses have the same forms as that of the vector field
because we started from a supersymmetric theory. We thus have a model
with $c_i=0$ for the spinors and $\bar c_i'=\bar\gamma'=0$ for the
scalar. (Note that $C_i$, which originate from $Q_i$, have $Z_2$ odd
parity.)

It is a nontrivial check to see whether the above mass terms satisfy
the conditions for 5D AdS$_5$ supersymmetry. We find from the
relations (\ref{eq:ci-spi}) and (\ref{eq:massd}) that the mass
terms for $\chi$ and $C$ imply
\begin{equation}
a \,=\, -4,\quad b \,=\, 2,\quad c \,=\, \frac{1}{2}.
\end{equation}
Indeed, these relations are just those required by AdS$_5$
supersymmetry~\cite{Gherghetta:2000qt}. 
In this way, {\it 5D vector supermultiplets on a RS warped
background are automatically derived from a 4D supersymmetric model 
on a flat background}.

\medskip

We also construct a 5D hypermultiplet in the warped extra dimension 
starting from a 4D supersymmetric theory. In order to have a 
hypermultiplet we introduce the chiral superfields $\phi_i$ 
and $\varphi_i$ in the fundamental and antifundamental representations 
of the ${\rm SU}(n)_i$ gauge theory. In the following, $\varphi_N$ is 
removed to implement $Z_2$ orbifolding which leaves a chiral zero 
mode of the fundamental representation. The fermionic components of 
$\phi_i$ and $\varphi_i$, then correspond to $\eta_i$ and $\psi_i$, 
respectively, in Sec.~\ref{sec:spinor}. The generic renormalizable 
superpotential is written as 
\begin{eqnarray}
W \,=\, \sum_{i=1}^{N-1}
(a_i \varphi^i Q_i \phi^{i+1} -b_iv_i \varphi^i \phi^i).
\label{eq:fspotential}
\end{eqnarray}
This superpotential just leads to a spinor mass term of the form
(\ref{eq:fermionmoose}). In addition, the mass and mixing terms of 
the scalars $\phi$ and $\varphi$ also have the same form as those 
of the spinors: 
\begin{eqnarray}
&& -\sum_{i=1}^{N-1} |a_iQ_i\phi_{i+1}+b_iv_i\phi_i|^2
  -\sum_{i=1}^{N-1} |a_{i-1}Q_{i-1}\varphi_{i-1}+b_iv_i\varphi_i|^2.
\end{eqnarray}
Supersymmetry induces equivalence between the boson and fermion 
mass matrices. In turn, this implies in our formulation given in the 
previous sections that the mass parameters are 
equal, $c_i=c'_i=\bar c'_i$ and also $\gamma_i=\bar\gamma_i=0$. Thus, 
there is only one parameter $c$ left. It is 
found from Eqs.~(\ref{eq:ci-spi}), (\ref{eq:massn}), and (\ref{eq:massd}) 
that if one take the continuum limit the relations
\begin{eqnarray}
&& a \,=\, \left(c+\frac{1}{2}\right)^2-4,\qquad b\,=\,\frac{3}{2}-c 
\qquad ({\rm for} \; \phi), 
\label{eq:abchypn} \\
&& a \,=\, \left(c-\frac{1}{2}\right)^2-4,\qquad b\,=\,\frac{3}{2}+c 
\qquad ({\rm for} \; \varphi) 
\label{eq:abchypd}
\end{eqnarray}
are generated. These mass relations are exactly those imposed by
supersymmetry on the AdS$_5$ geometry~\cite{Gherghetta:2000qt}. Thus
hypermultiplets on the RS background are properly incorporated in our
formalism with a flat background. It may be interesting that the
mass relation for vector multiplets is the one for chiral multiplets
with Dirichlet boundary conditions (\ref{eq:abchypd}) 
with $c=1/2$. This value of $c$ is the limit of vanishing bulk mass
parameters. 

It should be noticed that our analyses have been performed for generic
warped backgrounds, including the RS case as a special limit. We thus
found that even in generic warped backgrounds the conditions on 
the bulk mass parameters required for 5D warped supersymmetry should be
the same as for the RS case. 

\section{Numerical evaluation}
\label{sec:NE}

Here we perform a numerical study to confirm our formulation of the 
curved extra dimension discussed in the previous sections. We will 
also give a phenomenological application to the hierarchy among Yukawa 
couplings. 

\subsection{Spectrum and wave function}

In the following, we consider the case that corresponds to the RS
model in the continuum limit, as a good and simple application. The
gauge couplings and VEVs are specified as given in
Sec.~\ref{sec:DWD};
\begin{eqnarray}
\mbox{RS} : \qquad g_i \,=\, g, \qquad v_i \,=\, ve^{-ki/(gv)}.
\end{eqnarray}
The universal gauge coupling $^\forall g_i=g$ implies that vector
zero modes have flat wave functions as shown in Eq.~(\ref{zero-A}).
The following is a summary of the mass terms for various spin fields,
which were derived in the previous sections: 
\begin{eqnarray}
{\cal L}_{gm} &=& \frac{1}{2}\,\bigl|D_{1/2} A_\mu\bigr|^2, \\[1mm]
{\cal L}_{fm} &=& -\psi D_c \eta +\textrm{H.c.}, \\[2mm]
{\cal L}_{sm}^\phi &=& -\bigl|D_{3/2-b} \phi\bigr|^2 -|M \phi|^2,
\\[1mm]
{\cal L}_{sm}^\varphi &=& -\Bigl|D_{-(3/2-b)}^\dagger \varphi\Bigr|^2
-|M^\dagger \varphi|^2.
\end{eqnarray}
The parameters $b$ and $c$ represent the bulk mass parameters for scalars
and spinors, respectively. The $(N-1)\times N$ mass matrices $D_x$ and
$M$ are defined as follows:
\begin{eqnarray}
  D_x &=& g\pmatrix{ v_1 & & \cr & \ddots & \cr & & v_{N-1} }
  \pmatrix{ -1+\omega_x & 1 & & \cr
    & \ddots & \ddots & \cr
    & & -1+\omega_x & 1 },\\[2mm]
  M &=& \sqrt{a+4b-b^2}\,\frac{k}{v}
  \pmatrix{ v_1 & & \cr & \ddots & \cr & & v_{N-1} }
  \pmatrix{ 1 &    0   &         &   \cr
    & \ddots & \ddots  &   \cr
    &        & 1 & 0 },
\end{eqnarray}
where
\begin{equation}
\omega_x \,=\, \biggl(x-\frac{1}{2}\biggr)\frac{k}{gv}.
\end{equation}
For supersymmetric cases, the mass matrices $D_x$ for bosons and
fermions take the same form and, moreover, $M=0$, as discussed previously.

We define the matrices $U_{g,f,s}$ that diagonalize the mass matrices
for gauge, fermion, and scalar fields, respectively. For example, $U_g$
satisfies
\begin{eqnarray}
U_g^\dagger D_{1/2}^T D_{1/2} U_g \,=\, \textrm{diag}
\bigl((m_0^g)^2,(m_1^g)^2,\ldots,(m_{N-1}^g)^2\bigr),
\end{eqnarray}
where $m_i^g$ are the mass eigenvalues which should correspond to the 
KK spectrum of vector fields. In the following, we use the notation 
\begin{eqnarray}
U_j^g(i) &\equiv&  (U_g)^i_{\ j+1}, \quad 
i=1,\ldots,N, \ j=0,\ldots,N-1
\end{eqnarray}
that is, the coefficients of $A^i_\mu$ in the $j$th massive 
eigenstates $\tilde A^j_\mu$. In the continuum limit, this corresponds 
to the value of the wave function at $y=i/gv$ for the $j$th KK excited 
vector field. Similar definitions are made for spinors and scalars. 

\medskip

\begin{figure}[htbp]
\centerline{\epsfig{figure=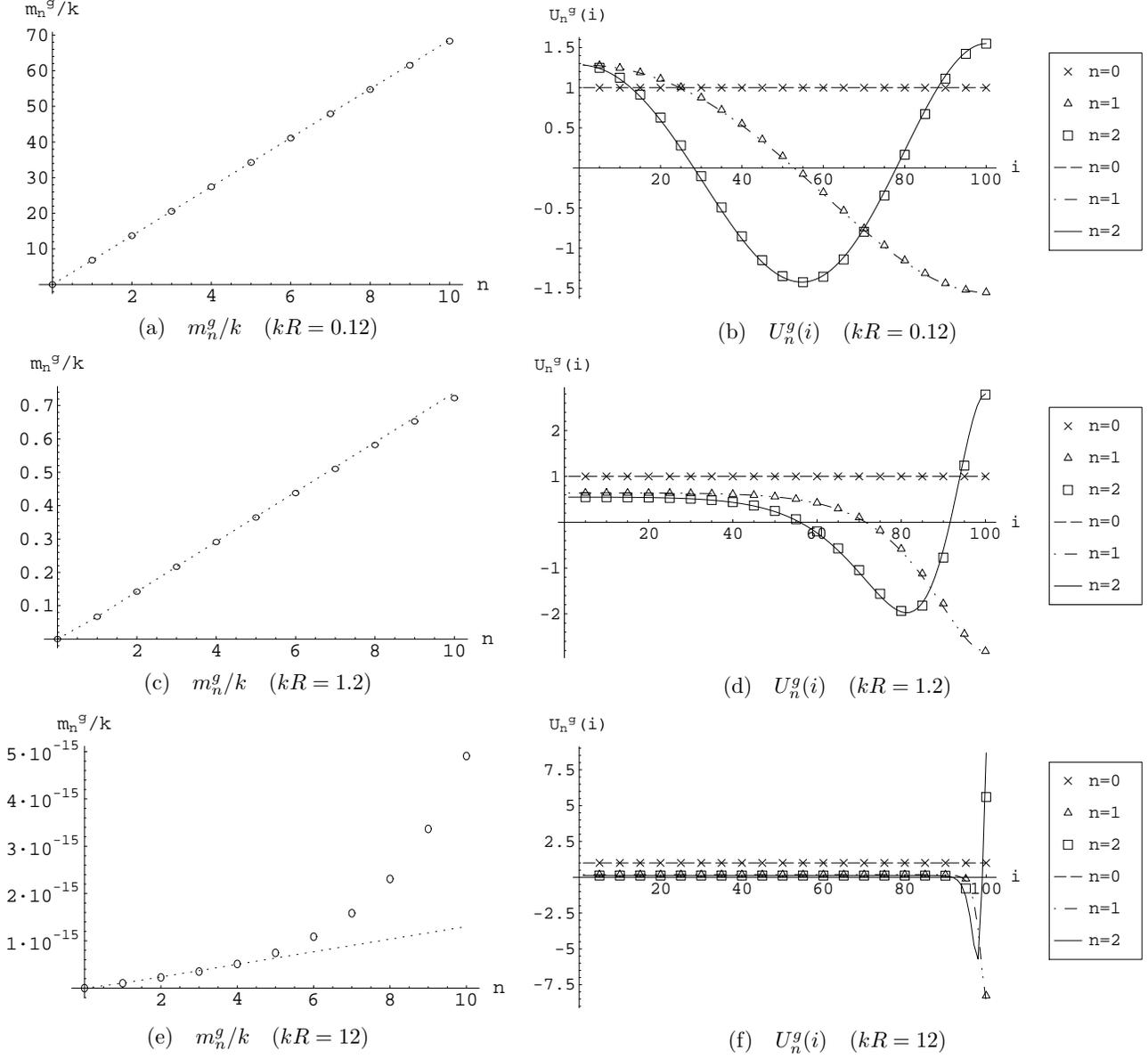,width=\linewidth}}
\caption{The mass eigenvalues and eigenvectors of the matrix $D_{1/2}$
  for vector fields in the cases $\,kL/2\pi=$ 0.12, 1.2 and 12. We
  take the total number of gauge groups as $N=100$. (a), (c), and
  (e) show the eigenvalues $m_n^g$ for each $kL$ with the 
  symbols $\circ$. The corresponding KK mass spectrum of the continuum
  RS theory is also depicted by the dotted lines. (b), (d), and
  (f) show the eigenvectors $U^g_n(i)$ for each $kL$ with the 
  symbols $\times$, $\triangle$, and $\Box$. The corresponding KK wave
  function of the continuum RS theory is also depicted by the
  lines. For the continuum cases, the horizontal axis is $gvy$. 
  The wave functions plotted here are normalized by the
  zero-mode ones.}
\label{fig:bulkgaugesys}
\end{figure}

For vector fields, we illustrate the resultant eigenvalues $m_n^g$ and 
eigenvectors $U^g_n(i)$ in 
Figs.~\ref{fig:bulkgaugesys}(a)--\ref{fig:bulkgaugesys}(f). 
For comparison, we also show in the figures the wave functions and KK mass
eigenvalues of vector fields on the RS background. It is found from
the figures that our 4D model completely reproduces the mode function
profiles [Figs.~\ref{fig:bulkgaugesys}(b), 
\ref{fig:bulkgaugesys}(d), and \ref{fig:bulkgaugesys}(f)]. Localization
becomes sharp as $kL$ increases; this situation is similar to the
continuum case. The warp-suppressed spectra of KK excited modes are
also realized [Figs.~\ref{fig:bulkgaugesys}(a), 
\ref{fig:bulkgaugesys}(c), and \ref{fig:bulkgaugesys}(e)]. 
For a larger $N$ (the number of gauge groups), the model leads to
a spectrum more in agreement with the continuous RS case. Note,
however, that the localization profiles of wave functions can be seen
even with a rather small $N$. It is interesting that even with a finite
number of gauge groups the massive modes have warp-suppressed
spectra and localization profiles in the index space of gauge
theory.

\begin{figure}[htbp]
\begin{center}
\begin{minipage}{0.7\linewidth}
\centerline{\epsfig{figure=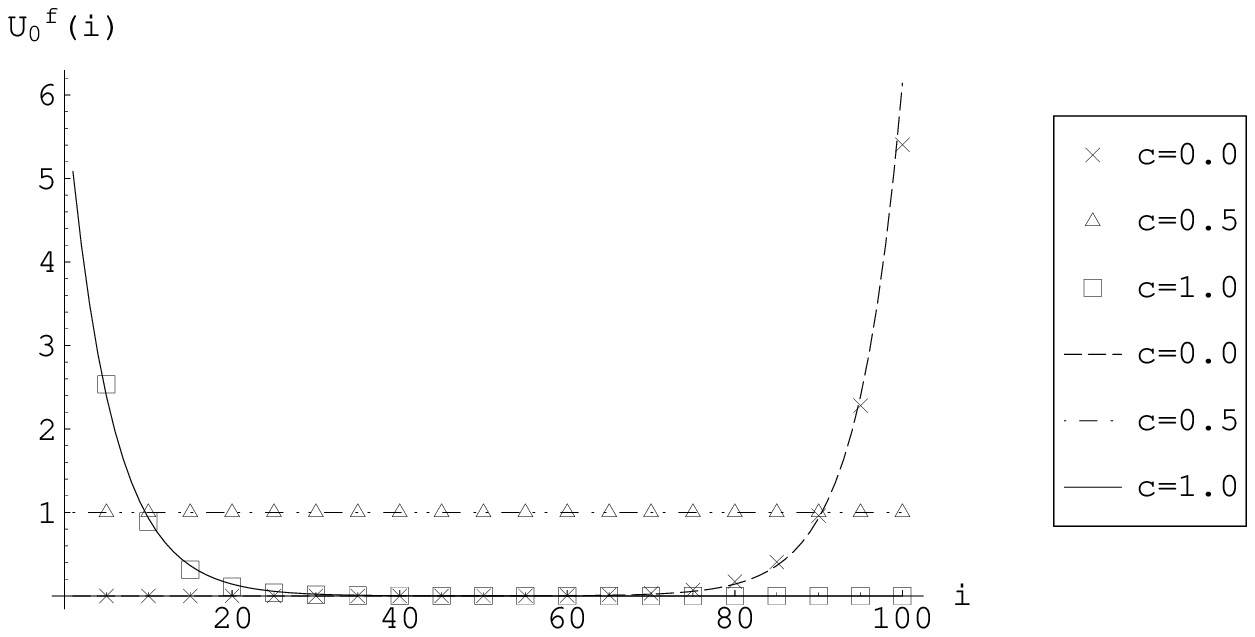,width=\linewidth}}
\end{minipage}
\end{center}
\caption{Typical behavior of the massless eigenvector $U^f_0(i)$ for
   fermions with mass parameter $c$ (denoted by the symbols
   $\times$, $\triangle$, and $\Box$). The total number of gauge groups
   is taken as $N=100$. The corresponding zero-mode wave functions in
   the continuum RS theory are depicted by the lines. The case $c=1/2$
   is the conformal limit where the massless mode is not localized.}
   \label{fig:cdepWF}
\end{figure}

For fermion fields, there is another interesting issue to be
examined. It is the localization behavior via dependence on the mass
parameters $c$, which was discussed in Sec.\ref{sec:spinor}. We
show the $c$ dependence of the zero-mode wave function $U^f_0(i)$ in
Fig.~\ref{fig:cdepWF}.
The figure indicates that the zero-mode wave functions surely give the
expected localization nature of the continuum RS limit
[Eq.~(\ref{eq:hzero}) in the Appendix]. We find that the values of the 
wave functions are exponentially suppressed at the tail of localization
profile even with a finite number of gauge groups. The profiles of
massive modes can also be reproduced.

\subsection{Yukawa hierarchy from 4D}

Now we apply our formulation to phenomenological problems in four 
dimensions. Let us use the localization behavior, which has been shown 
above, to obtain the Yukawa hierarchy. This issue has been studied in 
the 5D RS framework~\cite{Gherghetta:2000qt,Huber:2001ie}. We consider 
a model corresponding to the (supersymmetric) standard model in the 
bulk. The Yukawa couplings for quarks are given by 
\begin{eqnarray}
{\cal L}_{\rm Yukawa}
&=& \sum_{i=1}^N
    \left((y_u^{ab})_i\, q^i_a H^i_u u^i_b
    +(y_d^{ab})_i\, q^i_a H^i_d d^i_b \right)
    +\textrm{H.c.},
\label{eq:yukawaorg}
\end{eqnarray}
where $q$, $u$, and $d$ denote the left-handed quarks and the right-handed 
up and down quarks, respectively, and $a,b$ are the family
indices. For simplicity, we study a supersymmetric case and introduce
two types of Higgs scalars $H^i_u$ and $H^i_d$. Then the mass
parameters of the Higgs scalars satisfy Eq.~(\ref{eq:abchypn}) and
they are denoted by $c^{H_{u,d}}$ in the following. Similarly, the
quark behaviors are described by their mass parameters $c^{q,u,d}$. We
assume $(y_{u,d}^{ab})_i \sim {\cal O}(1)$. Generally, in
supersymmetric 5D models, Yukawa couplings such as Eq.~(\ref{eq:yukawaorg}) 
are prohibited by 5D supersymmetry. However, since the present model is 4D, 
one may apply 5D-like results to Yukawa couplings without respecting 5D 
consistency. This is one of the benefits of our scheme.

\begin{figure}[t]
\begin{center}
\begin{minipage}{0.45\linewidth}
\centerline{\hspace*{-5mm}\epsfig{figure=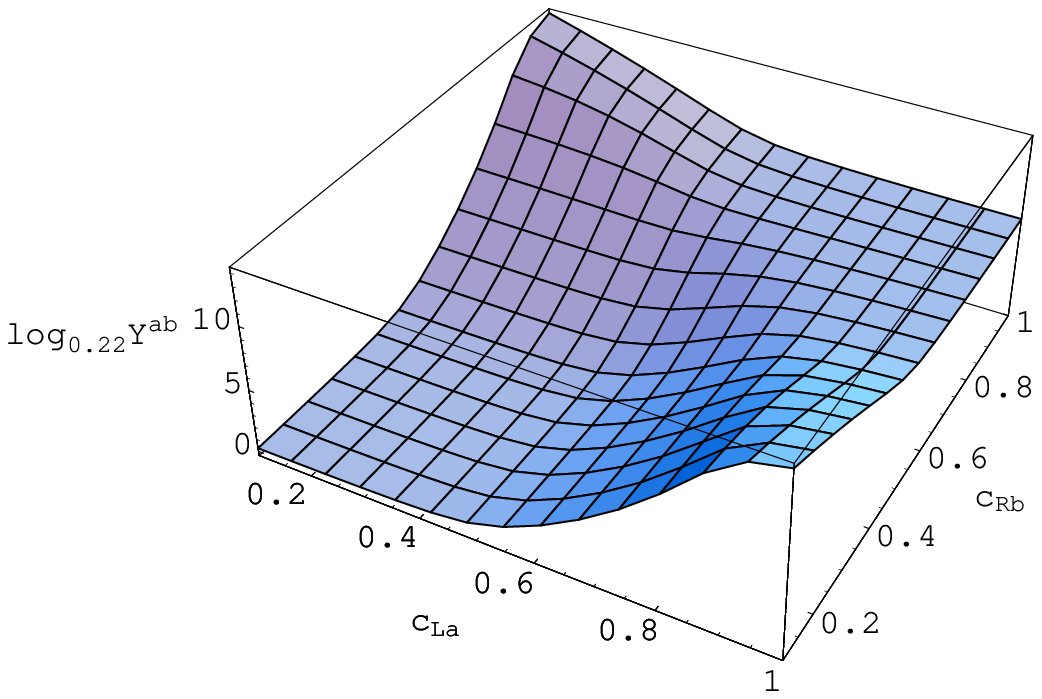,width=\linewidth}}
\vspace{-0.5cm}
\centerline{(a) $c^{H_{u,d}}=0$}
\end{minipage}
\hspace*{5mm}
\begin{minipage}{0.45\linewidth}
\centerline{\hspace*{-5mm}\epsfig{figure=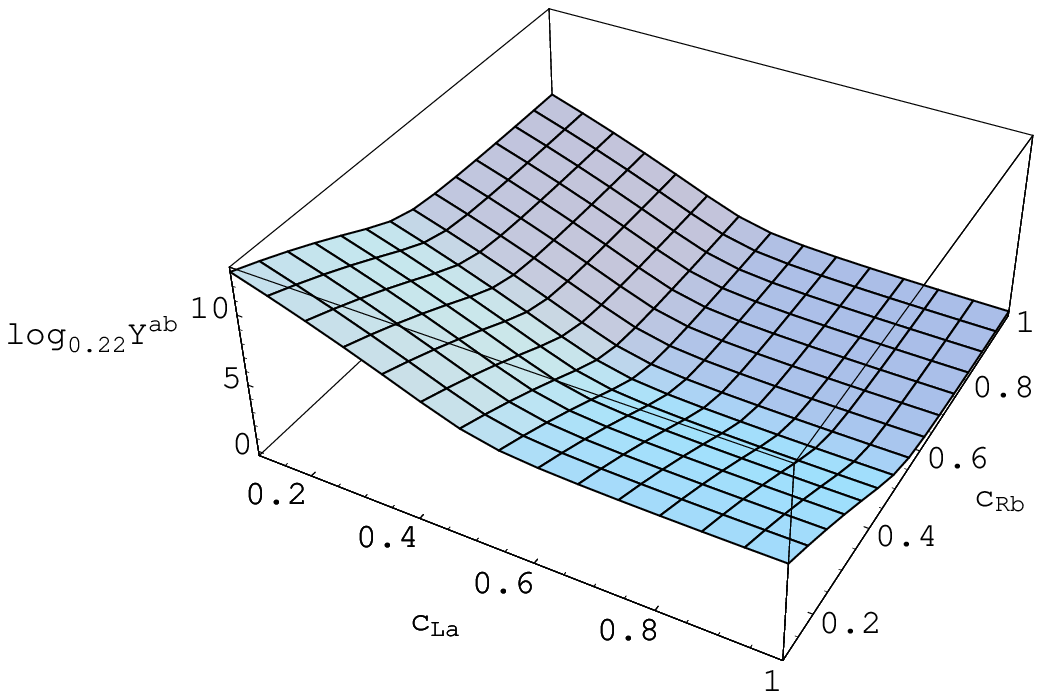,width=\linewidth}}
\vspace{-0.5cm}
\centerline{(b) $c^{H_{u,d}}=1$}
\end{minipage}
\begin{minipage}{0.45\linewidth}
\centerline{\hspace*{5mm}\epsfig{figure=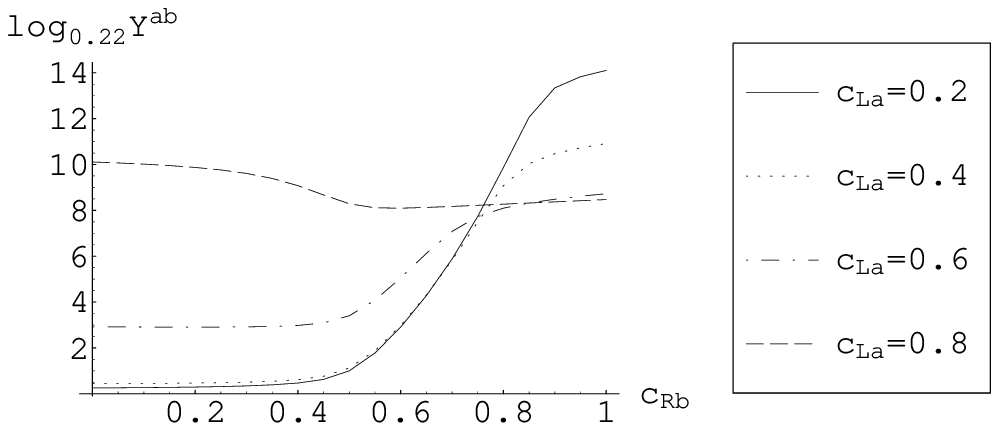,width=\linewidth}}
\centerline{(c) $c^{H_{u,d}}=0$}
\end{minipage}
\hspace*{5mm}
\begin{minipage}{0.45\linewidth}
\centerline{\hspace*{5mm}\epsfig{figure=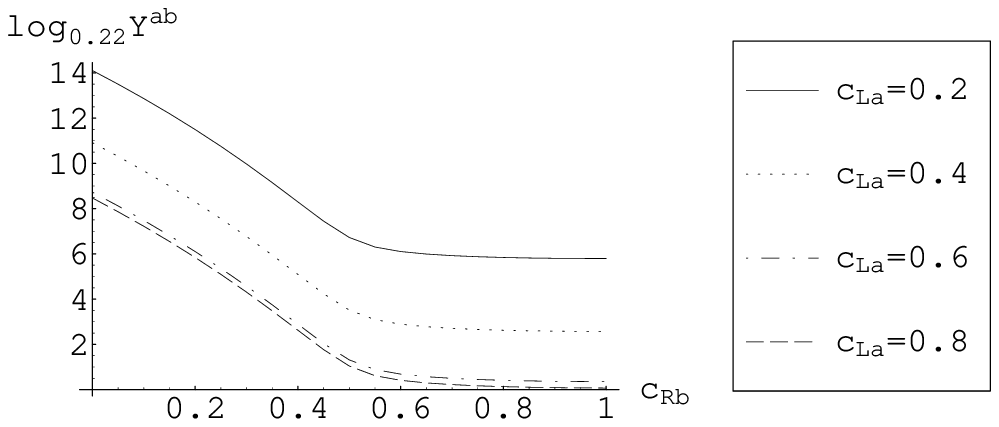,width=\linewidth}}
\centerline{(d) $c^{H_{u,d}}=1$}
\end{minipage}
\end{center}
\caption{The quark mass $c^q$ ($\equiv c_L$) 
  and $c^{u,d}$ ($\equiv c_R$) dependences of the zero-mode Yukawa
  coupling $Y^{ab}$ with $kL/2\pi=10.83$ and $N=20$. The mass
  parameters of the Higgs scalars $c^{H_{u,d}}$ are taken to be $0$ in
  (a) and (c), and $1$ in (b) and (d).}
\label{fig:cdepY}
\end{figure}

We are now interested in the zero-mode part of
Eq.~(\ref{eq:yukawaorg}), which generates the following mass terms 
\begin{eqnarray}
{\cal L}_{\rm Yukawa}
&=& Y_u^{ab} \tilde{q}^0_a \langle \tilde{H}^0_u \rangle \tilde{u}^0_b
   +Y_d^{ab} \tilde{q}^0_a \langle \tilde{H}^0_d \rangle \tilde{d}^0_b
   \,+\textrm{H.c.},
\end{eqnarray}
where the fields with tildes $\tilde{q}^j$ stand for the $j$th mass
eigenstate given by $\tilde{q}^j = \sum_{i=1}^N U_j^q(i)\, q^i$
(similarly for $u$, $d$, and $H_{u,d}$). The effective Yukawa couplings
are
\begin{eqnarray}
Y_u^{ab} &=& y_u^{ab}
\sum_{i=1}^N \,U_0^{H_u}(i)\,U_0^{q_a}(i)\,U_0^{u_b}(i),
\end{eqnarray}
and similarly for $Y_d$. A typical behavior of $U_0(i)$ is shown in
Fig.~\ref{fig:cdepWF} for several values of the bulk mass parameter $c$.
In Fig.~\ref{fig:cdepY}, we show the behaviors of the zero-mode Yukawa
couplings $Y_{u,d}$ against the quark mass parameters. Two limiting
cases with $c^{H_{u,d}}=0$ and 1 are shown. The former corresponds
to a bulk Higgs scalar localized at $y=L/2$ and the latter to one at $y=0$
in the continuum RS limit. From the figures, we see that if there is
a ${\cal O}(1)$ difference of mass ratio among the generations, it
generates a large hierarchy between Yukawa couplings. Combined 
with the mechanisms that control mass parameters discussed in the next
section, one obtains a hierarchy without symmetries within the 
four-dimensional framework.

In the case of $c^{H_{u,d}}=1$, the Yukawa coupling depends
exponentially on the quark bulk mass parameters $c^{q,u,d}$ 
when $c^{q,u,d}<0.5$.\footnote{Similar behavior can be obtained for
$c^{H_{u,d}} > 0.5$. For $c^{H_{u,d}} > 0.5$, the Higgs scalars
have a peak at $i=1$ ($y=0$). This situation is different from
the one discussed in Ref.~\cite{Huber:2001ie} where the Higgs field
is localized at $i=N$ ($y=L/2$).}
This implies that if $c^{q,u,d}$ exist in this region one obtains the
following form of the Yukawa matrices:
\begin{equation}
\pmatrix{\lambda^{n_{aa}} & \lambda^{n_{ab}} \cr
         \lambda^{n_{ba}} & \lambda^{n_{bb}} },
\label{yij}
\end{equation}
where their exponents satisfy
\begin{equation}
n_{aa}+n_{bb} \,=\, n_{ab}+n_{ba}.
\label{FNtype}
\end{equation}
This form is similar to that obtained by the Froggatt-Nielsen
mechanism~\cite{Froggatt:1979nt} with a ${\rm U}(1)$ symmetry. As an
illustration, let us take the following mass parameters
\begin{eqnarray}
c^{q_{1,2,3}} &=& (0.36,\ 0.42,\ 1.00), \nonumber \\
c^{u_{1,2,3}} &=& (0.36,\ 0.42,\ 1.00), \\
c^{d_{1,2,3}} &=& (0.42,\ 1.00,\ 1.00), \nonumber \\
c^{H^{u,d}}   &=& 1, \nonumber 
\end{eqnarray}
and $N=20$ and $kR=10.83$, which generates the low-energy Yukawa
matrices
\begin{eqnarray}
Y_u^{ab} \,\simeq\,
\pmatrix{\lambda^{6} & \lambda^{5} & \lambda^{3} \cr
         \lambda^{5} & \lambda^{4} & \lambda^{2} \cr
         \lambda^{3} & \lambda^{2} & 1 }, \qquad
Y_d^{ab} \,\simeq\,
\pmatrix{\lambda^{5} & \lambda^{3} & \lambda^{3} \cr
         \lambda^{4} & \lambda^{2} & \lambda^{2} \cr
         \lambda^{2} & 1 & 1 },
\end{eqnarray}
where $\lambda = 0.22$. This pattern of quark mass textures leads to
realistic quark masses and mixing angles~\cite{Altarelli:1999ns} with
a large value of the ratio
$\langle \tilde{H}_u^0\rangle / \langle \tilde{H}_d^0 \rangle$.
If the above analysis were extended to SU(5) grand unified theory,
realistic lepton masses and mixing may be derived. Other forms of
Yukawa matrices that may be realized by the Froggatt-Nielsen mechanism
are easily incorporated in our formulation.

For more complicated patterns of mass parameters, we could realize
Yukawa matrices that are different from those derived from the
Froggatt-Nielsen mechanism. In general, off-diagonal entries tend to
be rather suppressed, that is, we have
\begin{equation}
n_{aa}+n_{bb} \,\leq\, n_{ab}+n_{ba}
\end{equation}
for the Yukawa matrix (\ref{yij}). Such a form may lead to realistic
fermion masses and mixing angles. For example, one could derive the
Yukawa matrix 
\begin{eqnarray}
Y^{ab} \,\simeq\,
\pmatrix{ 0  & \lambda^{4} & 0 \cr
         \lambda^{4} & \lambda^{3} & \lambda^{3} \cr
         0 & \lambda^{3}  & 1 }
\label{eq:compli}
\end{eqnarray}
if initial values of $(y^{11})_i$ are sufficiently suppressed. In this
case, the $2 \times 2$ submatrix for the second and third generations
does not satisfy Eq.~(\ref{FNtype}). The Yukawa matrix
(\ref{eq:compli}) may be relevant to the down-quark sector, indeed
studied in Ref.~\cite{Ramond:1993kv}. We do not pursue further
systematic studies on these types of Yukawa matrices in this paper.

\section{Toward dynamical realization}
\label{sec:DDWD}

We have shown that 4D models with nonuniversal VEVs and gauge and
other couplings can describe 5D physics on curved backgrounds,
including the RS model with an exponential warp factor. In the
continuum 5D theory, this factor is derived as a solution of the
equation of motion for gravity. On the other hand, in the 4D
viewpoint, warped geometries are generated by taking the couplings and
VEVs as appropriate forms. In the previous sections, we have just
assumed their typical forms and examined its consequences. If one
could identify how to control these couplings by the underlying 
{\it dynamics}, the resultant 4D theories turn out to provide
attractive schemes to discuss low-energy physics such as tiny coupling
constants.

First we consider the scalar VEVs $\langle Q_i\rangle=v_i$.
A simple way to dynamically control them is to introduce additional
strongly coupled gauge theories~\cite{Arkani-Hamed:2001ca}. Consider
the following set of asymptotically free gauge theories:
\begin{eqnarray}
A_\mu^i &:& {\rm SU}(n)_i\textrm{ gauge field }(g,\Lambda), \\
\hat A_\mu^i &:& {\rm SU}(m)_i\textrm{ gauge field }
(\hat{g}_i,\hat{\Lambda}_i),
\end{eqnarray}
where $\Lambda$ and $\hat{\Lambda}_i$ denote the dynamical scales. We
have, for simplicity, assumed common values of $g$ and $\Lambda$
for all ${\rm SU}(n)_i$. In addition, two types of fermions are introduced: 
\begin{eqnarray}
\xi^i \;:\; (n,\bar{m},1), \qquad
\bar\xi^i \;:\; (1,m,\bar{n}),
\end{eqnarray}
where their representations 
under ${\rm SU}(n)_i \times {\rm SU}(m)_i \times {\rm SU}(n)_{i+1}$ gauge 
groups are shown in parentheses. If $\hat{\Lambda}_i \gg \Lambda$, 
the ${\rm SU}(m)_i$ theories are confined at a higher scale than 
${\rm SU}(n)_i$, and the fermion bilinear composite scalars 
$Q_i \sim \xi^i\bar\xi^i$ appear. Their VEVs $v_i$ are given by the 
dynamical scales $\hat{\Lambda}_i$ of the ${\rm SU}(m)_i$ gauge theories 
through a dimensional transmutation as
\begin{eqnarray}
\langle Q_i \rangle \,\equiv\, v_i \,\simeq\, \hat{\Lambda}_i
\,=\, \mu\, e^{1/2\hat{\beta} \hat{g}_i^2(\mu) },
\label{eq:DTvev}
\end{eqnarray}
where $\hat{\beta}$ is a universal one-loop gauge beta function 
for ${\rm SU}(m)_i$ ($\hat\beta<0$). The gauge couplings $\hat g_i$
generally take different values and thus lead to different values 
of $v_i$. For example, a linear dependence of $1/\hat g_i^2$ on the
index $i$ is amplified to an exponential behavior of $v_i$. That is,
\begin{eqnarray}
\frac{1}{\hat{g}_i^2(\mu)}
\,=\, -2\hat\beta\biggl(\frac{k}{gv}\biggr)i
\quad\longleftrightarrow\quad v_i \,=\, ve^{-ki/gv},
\end{eqnarray}
which reproduces the bulk fields on the RS background as shown before.
The index dependences of the gauge couplings are actually generic 
situations, and may also be controlled, for example, by some mechanism 
fixing dilatons or the radiatively induced kinetic terms discussed below.
A supersymmetric extension of the above scenario is achieved with
quantum-deformed moduli spaces~\cite{Csaki:2002em}.

\medskip

Another mechanism that dynamically induces nonuniversal VEVs is
obtained in supersymmetric cases. Consider the gauge
group $\prod_i {\rm U}(1)_i$ and the chiral superfields $Q_i$ with
charges $(+1,-1)$ under ${\rm U}(1)_i\times {\rm U}(1)_{i+1}$. 
It is assumed that the scalar components $q_i$ of $Q_i$ develop their
VEVs $\langle q_i \rangle = v_i$. The $D$ term of each ${\rm U}(1)_i$ 
is given by
\begin{equation}
D_i \,=\, \epsilon_i + |q_i|^2 - |q_{i-1}|^2  + \cdots,
\end{equation}
where $\epsilon_i$ is the coefficient of the Fayet-Iliopoulos (FI)
term, and the ellipsis denotes contributions from other fields 
charged under ${\rm U}(1)_i$, which are assumed not to have VEVs. 
Given nonvanishing FI terms, $\epsilon_i \neq 0$, the $D$-flatness 
conditions mean 
\begin{equation}
v_{i-1}^2 \,=\, v_i^2 + \epsilon_i,
\end{equation}
and nonuniversal VEVs $v_i$ are indeed realized. In this case, the 
dynamical origin of nonuniversal VEVs is the nonvanishing FI 
terms. These may be generated at the loop level. Furthermore, if the 
matter content is different for each gauge theory, the $\epsilon_i$ 
themselves have complicated forms.

Above, we supposed that the charges of $Q_i$ are $(+1,-1)$ 
under ${\rm U}(1)_i \times {\rm U}(1)_{i+1}$. Alternatively, if $Q_i$ 
have charges $(M,-M)$ under ${\rm U}(1)_i \times {\rm U}(1)_{i+1}$ and 
other matter fields have integer charges, the gauge symmetry 
$\prod_i {\rm U}(1)_i$ is broken to the product of a diagonal 
${\rm U}(1)$ gauge symmetry and the discrete gauge symmetry 
$\prod_i(Z_M)_i$. Such discrete gauge symmetry
would be useful for phenomenology~\cite{Ibanez:1991pr}.

\medskip

Models with nonuniversal gauge couplings $g_i$ are also interesting
in the sense that they can describe the localization of massless
vector fields. A nonuniversality of gauge couplings is generated,
e.g.,\ in the case that the ${\rm SU}(n)_i$ gauge theories have different
matter content from each other. Then radiative corrections to gauge
couplings and their renormalization-group running become
nonuniversal, even if their initial values are equal.

This fact is also applicable to the above-mentioned mechanism for
nonuniversal $v_i$. Suppose that the ${\rm SU}(m)_i$ theory contains
($2m+i$) vectorlike quarks which decouple at $v$. The gauge
couplings $\hat g_i(v)$ are then determined by
\begin{equation}
 \frac{1}{\hat g_i^2(v)} \;=\; \frac{i}{8\pi^2}
 \ln\biggl(\frac{\Lambda'}{v}\biggr),
\label{eq:nonuni-g}
\end{equation}
where we have assumed that the ${\rm SU}(m)_i$ theories are strongly 
coupled at a high-energy scale $\Lambda'$~($>v$). Tuning of the relevant 
matter content thus generates the desired linear dependence 
of $1/\hat g_i^2$. With these radiatively induced couplings
(\ref{eq:nonuni-g}) at hand, the VEVs are determined from
Eq.~(\ref{eq:DTvev}): 
\begin{equation}
 v_i \;=\; v\biggl(\frac{v}{\Lambda'}\biggr)^{i/2m}.
\end{equation}

\section{Conclusion}
\label{sec:conclusion}

We have formulated 4D models that provide 5D field theories on
generic warped backgrounds. The warped geometries are achieved with
generic values of symmetry-breaking VEVs, gauge couplings, and other
couplings in the models. We focused on field localization
behaviors along the index space of gauge theory (the fifth dimension
in the continuum limit), which is realized by taking relevant choices
of the mass parameters.

As a good and simple application, we constructed 4D models
corresponding to bulk field theories on the AdS$_5$ Randall-Sundrum
background. The localized wave functions of massless modes are
completely reproduced with a finite number of gauge groups.
In addition, the exponentially suppressed spectrum of the KK modes is
also generated. These results imply that most properties of brane
world models can be obtained within 4D gauge theories. Supersymmetric
extensions were also investigated. In 5D warped models, the bulk and
boundary mass terms of spinors and scalars satisfy complicated forms
imposed by supersymmetry on the RS background. However, we show in our
formalism that these forms of the mass terms are derived from a 
4D {\it global} supersymmetric model on a {\it flat} background.

As an application of our 4D formulation, we derived hierarchical
forms of Yukawa couplings. The zero modes of scalars and spinors with
different masses have different wave-function profiles as in the
5D RS cases. Therefore by varying the ${\cal O}(1)$ mass parameters for
each generation, one can obtain realistic Yukawa matrices with a large
hierarchy from the overlaps of the wave functions in a purely 4D
framework. Other phenomenological issues such as proton stability,
grand unified theory (GUT) symmetry breaking, and supersymmetry 
breaking can also be discussed.

The conditions on the model parameters should be explained by some
dynamical mechanisms if one considers the models from a fully 4D
viewpoint. One interesting way is to include additional
strongly coupled gauge theories. In this case, a 
small ${\cal O}(1)$ difference between gauge couplings is converted to
exponential profiles of symmetry-breaking VEVs via dimensional
transmutation, and indeed generates a warp factor of the RS model.
A difference of gauge couplings is achieved by, for example, the dynamics
controlling dilatons, or radiative corrections to gauge couplings.
Supersymmetrizing models provide a mechanism for dynamically realizing
nonuniversal VEVs with $D$-flatness conditions.

Our formulation makes sense not only from the 4D points of view but
also as a lattice-regularized 5D theory. In this sense, effects such 
as the AdS/conformal field theory (CFT) correspondence might be clearly 
seen with our formalism. As another application, it can be applied to 
construct various types of curved backgrounds and bulk or boundary 
masses. For example, we discussed massless vector localization by varying 
the gauge couplings $g_i$. Furthermore, one might consider models in 
which some fields are charged under only some of the gauge groups. 
These seem not like bulk or brane fields, but ``quasi-bulk'' 
fields. Applications including these phenomena will be studied
elsewhere.

\subsection*{Acknowledgment}
This work is supported in part by the Japan Society for
the Promotion of Science under the Postdoctoral Research Program
(Grants No.~$08557$ and No.~$07864$) and a Grant-in-Aid for Scientific 
Research from the Ministry of Education, Science, Sports and Culture 
of Japan (No.~$14540256$).

\bigskip

\section*{Appendix. ~Bulk fields in AdS$_5$}

Here we briefly review the field theory on a RS background,
following Ref.~\cite{Gherghetta:2000qt}. One of the original
motivations for introducing a warped extra dimension by Randall and
Sundrum is to provide the weak Planck mass hierarchy via the
exponential factor in the space-time metric. This factor is called
``warp factor,'' and the bulk space a ``warped extra dimension''. 
Such a nonfactorizable geometry with a warp factor distinguishes 
the RS brane world from others.

Consider the fifth dimension $y$ compactified on an orbifold
$S^1/{\bf Z}_2$ with radius $R$ and two three-branes at the orbifold
fixed points $y=0$ and $y=L/2 \equiv \pi R$. 
The Einstein equation for this five-dimensional setup leads to 
the solution~\cite{Randall:1999ee} 
\begin{equation}
ds^2 \,=\, e^{-2\sigma}\eta_{\mu\nu}dx^\mu dx^\nu-dy^2,
\qquad \sigma \,=\, k|y|,
\label{metric}
\end{equation}
where $k$ is a constant with mass dimension $1$. Let us study a vector
field $A_M$, a Dirac fermion $\Psi$, and a complex scalar $\phi$ in the
bulk specified by the background metric (\ref{metric}). The 5D action
is given by
\begin{equation}
  S_5 \,=\, \int d^4x\int_0^{\pi R} dy\sqrt{-g}\,
    \Bigg[\frac{-1}{4g^2_5}F^2_{MN}+
    \left|\partial_M\phi\right|^2+i\bar{\Psi}\gamma^MD_M\Psi
    -m^2_\phi|\phi|^2-m_\Psi\bar{\Psi}\Psi\Bigg],
\label{kin}
\end{equation}
where $\gamma_M=(\gamma_\mu,i\gamma_5)$ and the covariant derivative
is $D_M=\partial_M+\Gamma_M$ where $\Gamma_M$ is the spin connection
given by $\Gamma_\mu= i\gamma_5\gamma_\mu \sigma'/2$ 
and $\Gamma_4=0$. From the transformation properties under 
${\bf Z}_2$ parity, the mass parameters of scalar and fermion
fields are parametrized as\footnote{In Ref.~\cite{Gherghetta:2000qt},
the integral range with respect to $y$ is taken
as $-\pi R\leq y\leq\pi R$. Here we adopt $0\leq y\leq \pi R$, and
then the boundary mass parameter $b$ in Eq.~(\ref{eq:smass}) is
different from that in Ref.~\cite{Gherghetta:2000qt}
by the factor $1/2$.}
\begin{eqnarray}
m^2_\phi &=& ak^2+\frac{b}{2}\sigma'',\label{eq:smass} \\
m_\Psi &=& c\sigma' \label{eq:fmass},
\end{eqnarray}
where $a$, $b$, and $c$ are dimensionless parameters.

Referring to \cite{Gherghetta:2000qt}, the vector, scalar and spinor
fields are cited together using the single notation
$\Phi=\{A_\mu,\ \phi,\ e^{-2\sigma}\Psi_{L,R}\}$.
The KK mode expansion is performed as
\begin{equation}
\Phi(x^\mu,y) \,=\, {1\over\sqrt{2\pi R}}
\sum_{n=0}^\infty \Phi^{(n)}(x^\mu)f_n(y).
\end{equation}
By solving the equations of motion, the eigenfunction $f_n$ is given
by
\begin{eqnarray}
f_n(y)
&=& \frac{e^{s\sigma/2}}{N_n}
    \left[J_\alpha \left(\frac{m_n}{k}e^{\sigma} \right)
   +b_{\alpha}(m_n)\, 
          Y_\alpha \left(\frac{m_n}{k}e^{\sigma} \right) \right],
  \label{eq:modefunction}
\end{eqnarray}
where $\alpha = \sqrt{(s/2)^2+M^2_\Phi/k^2}$,
$s=\{2,\ 4,\ 1\}$, and $M^2_\Phi=\{0,\ ak^2,\ c(c\,\pm \, 1)k^2\}$
for each component in $\Phi$. $N_n$ is the normalization factor 
and $J_\alpha$ and $Y_\alpha$ are the Bessel functions. The
corresponding KK spectrum $m_n$ is obtained by solving
\begin{eqnarray}
b_{\alpha}(m_n) \,=\, b_{\alpha}(m_ne^{\pi kR}).
\label{eq:KKequation}
\end{eqnarray}

A supersymmetric extension of this scenario was discussed
in~\cite{Gherghetta:2000qt,Gherghetta:2001kr}. The on-shell field
content of a vector supermultiplet is $(A_M,\lambda_i,\Sigma)$ where
$A_M$, $\lambda_i~(i=1,2)$, and $\Sigma$ are the vector, two Majorana
spinors, and a real scalar in the adjoint representation,
respectively. Also a hypermultiplet consists of $(H_i,\Psi)$, 
where $H_i$ $(i=1,2)$ are two complex scalars and $\Psi$ is a Dirac
fermion. By requiring the action (\ref{kin}) to be invariant under
supersymmetric transformation on the warped background, one finds that
the five-dimensional masses of the scalar and spinor fields have to
satisfy
\begin{eqnarray}
m^2_\Sigma&=&-4k^2+2\sigma'', \label{eq:Sig} \\
m_\lambda&=&\frac{1}{2}\sigma', \label{eq:lam} \\
m^2_{H^{1,2}}&=&\left(c^2\pm c-\frac{15}{4}\right)k^2
 +\left(\frac{3}{2}\mp c\right)\sigma'',\label{eq:H} \\
m_\Psi&=&c\sigma', \label{eq:Psi}
\end{eqnarray}
where $c$ remains as an arbitrary dimensionless parameter.
That is, $a=-4$, $b=2$, and $c=1/2$ for vector multiplets and
$a=c^2\pm c-15/4$ and $b=3/2 \mp c$ for hypermultiplets. There is no
freedom to choose the bulk masses for vector supermultiplets and
only one freedom parametrized by $c$ for the bulk hypermultiplets.
It should be noted that in warped 5D models fields contained in
the same supermultiplet have different bulk and boundary masses.
That is in contrast with the flat case.

The $Z_2$ even components in supermultiplets have massless modes
with the following wave functions: 
\begin{eqnarray}
\frac{1}{\sqrt{2\pi R}} ~~&&
\textrm{ for } \;V^{(0)}_\mu \,
\textrm{ and } \,\lambda^{1\,(0)}_L, \\[1mm]
\frac{e^{(1/2- c)\sigma}}{N_0\sqrt{2\pi R}} &&
\textrm{ for } \;H^{1\, (0)} \,
\textrm{ and } \,\Psi^{(0)}_L. \label{eq:hzero}
\end{eqnarray}
The subscript $L$ means the left-handed ($Z_2$ even) component.
The massless vector multiplet has a flat wave function in the extra
dimension. On the other hand, the wave function for massless chiral
multiplets involves a $y$-dependent contribution from the space-time
metric, which induces a localization of the zero modes. The zero modes
with masses $c>1/2$ and $c<1/2$ localize at $y=0$ and $y=\pi R$,
respectively. The case with $c=1/2$ corresponds to the conformal limit
where the kinetic terms of the zero modes are independent of $y$.

\end{document}